\documentclass[conference]{IEEEtran}
\usepackage{graphicx} 
\usepackage{soul}
\title{A Taxonomy of Human-Robot Teamwork Requirements}

\author{Anastasia Mavridou, Hazel M. Taylor, Sandy Lozito, Louise Dennis, Michael Fisher, Marie Farrell}

\author{
  Anastasia Mavridou$^{1*}$, Hazel M. Taylor$^{2*}$, Sandy Lozito$^3$,
  Louise A. Dennis$^2$, Michael Fisher$^2$, Marie Farrell$^2$\\
  $^1$KBR Inc. at NASA Ames Research Center, Moffett Field, USA\\
  $^2$University of Manchester, Manchester, UK\\
  $^3$NASA Ames Research Center, Moffett Field, USA 
}

\usepackage{xcolor}

\newcommand{\category}[1]{{\small\textsc{{#1}}}}
\newcommand{\new}[1]{\textcolor{black}{#1}}

\usepackage{paralist}
\usepackage{url}
\usepackage{booktabs}
\usepackage{multirow}
\usepackage{graphicx}
\usepackage{array}
\usepackage[table]{xcolor}

\usepackage{todonotes}

\begin{document}

\maketitle

\begin{abstract}
    Autonomous systems are increasingly deployed in safety- and mission-critical domains where humans and robots must operate as a team to complete complex tasks. Existing requirements for Human-Robot teamwork remain fragmented across disparate sources, with no unified framework that addresses complexities of collaborative Human-Robot tasks. We address this gap by presenting a taxonomy of Human-Robot Teamwork (HRT) requirements derived from analysis of (academic and industrial) literature, standards and regulatory guidance. We extracted \new{a construction corpus of }$361$ requirements from $14$ cross-domain sources.
    Through iterative classification and refinement, we develop a two-level hierarchical taxonomy comprising $6$ high-level categories and $21$ low-level subcategories that distinguish information provision, relational control, decision support, safety mechanisms, performance monitoring, and foundational system capabilities. \new{We validate the taxonomy through expert evaluation with 5 domain specialists and a utility demonstration on an independently assembled corpus of $448$ requirements drawn from $19$ sources 
    spanning six HRT domains.} 
    \new{The taxonomy classifies $412$ of these requirements, with distributional patterns that converge with the construction corpus while revealing domain-specific patterns and gaps.} 
    We examine how requirements distribute across Human-Led, Robot-Led and Shared operational perspectives, revealing responsibility boundaries that shape safe collaboration. 
    This work provides a structured foundation for Requirements Engineering (RE) practitioners to systematically elicit and specify HRT capabilities.
    \def\thefootnote{*}\footnotetext{These authors contributed equally to this work.}\def\thefootnote{\arabic{footnote}}
\end{abstract}

\section{Introduction}
\label{sec:intro}

Autonomous systems are increasing in complexity and capability, enabling new possibilities for collaboration between humans and robots in critical settings. Human-Robot teams are increasingly being tasked with completing complex operations in a plethora of domains, including autonomous robotic systems that (1) reduce human exposure to hazardous environments (e.g., nuclear applications~\cite{fisher2021overview}), and (2) enable operations that are simply not possible for humans in isolation due to hazardous and hostile environments (e.g., space missions~\cite{etumi2025space}). In these critical settings, humans must work alongside AI-driven robots to complete tasks that require complementary capabilities, shared decision-making, and coordinated action.

Requirements are central to software development in safety-critical domains. Yet, for HRT, they remain fragmented across disparate standards, industry technical reports, and academic literature. 
This hinders the systematic deployment of Human-Robot teams in critical areas such as aerospace, nuclear operations, and advanced manufacturing, where comprehensive requirements are mandatory. 

This work fills this gap with a taxonomy and corpora for \new{HRT requirements} that organize distributed and disparate sources into a coherent hierarchy. It supports traceability and comparison across systems and team structures, providing requirements engineers with a reference against which specifications can be assessed for coverage and clarity. We seek to answer two research questions:





\begin{itemize}
    \item[\textbf{RQ1:}] What are suitable categories of HRT requirements?
    \item[\textbf{RQ2:}] To what extent can these categories capture diverse patterns of HRT?
\end{itemize}

For \textbf{RQ1}, we consider categories as \textit{suitable} if they are coherent and well-defined, cover Human-Robot teamwork requirement types without excessive overlap, and are judged useful by domain experts for elicitation and specification tasks.
We address \textbf{RQ1} by deriving a structured taxonomy from the sources listed in Table~\ref{tab:sources}. \new{For \textbf{RQ2}, we employ two complementary validation methods: 
1) expert evaluation with 5 domain experts; and 2) a utility demonstration~\cite{usman2017taxonomies} on an independently assembled validation corpus.}

\textbf{Contributions.} In this paper, we contribute:
\begin{itemize}
    \item \textbf{A \textit{\new{construction} corpus} of 361 requirements} for Human-Robot teams extracted from standards, regulatory documents and peer-reviewed literature (\S\ref{sec:methodology}).
    \item \textbf{A two-level hierarchical taxonomy} of requirements for Human-Robot teamwork comprising 6 high-level and 21 low-level categories, each characterized by distinct teamwork patterns 
    (\S\ref{sec:highleveltaxonomy} and \S\ref{sec:lowleveltaxonomy}).
    \item \textbf{An expert evaluation} 
    validates the taxonomy's conceptual clarity and explores Human-Led, Robot-Led and Shared operational perspectives (\S\ref{sec:expertevaluation}).
\new{\item \textbf{A \textit{validation corpus} of 448 requirements}  
across 6 HRT domains, used to validate the taxonomy with \textbf{a utility demonstration} that classifies 412 out-of-sample requirements and highlights domain differences (\S\ref{sec:utilitydemonstration}).}
\end{itemize}

\section{Background and Related Work }
\label{sec:related}



Human-Robot teams collaborate through \emph{supervisory control} (humans oversee and intervene), \emph{shared decision-making} (combining strengths to achieve goals), and \emph{independent collaboration} (performing distinct tasks while sharing information). Their growing use in critical domains demands verifiable requirements (
e.g., DO-178C~\cite{rierson2017developing}), yet current requirements focus on technical correctness while not adequately capturing teamwork aspects critical to reliability and safety.

A recent NASA technical standard \cite{nasa-std-3001} defines human-system interaction requirement classes spanning crew health, performance, training, and hardware/software design. We draw inspiration from \cite{nasa-std-3001} and reuse its nomenclature where applicable.
We also refine categories from \cite{nasa-std-3001} to more accurately reflect HRT and introduce new categories not present in \cite{nasa-std-3001}.  
The Brahms language provides semi-formal descriptions of HRT for space applications~\cite{ClanceySKH03, Brahms20:AERO}. 
Mission requirement patterns for autonomous mobile robots are proposed in~\cite{vazquez2024robotics}, though they do not capture the human. 
Explainability requirements and patterns are studied in \cite{droste2024explanations,refsq,icra}. Our taxonomy includes these but focuses more broadly on HRT, complementing existing work. \cite{hendry2025roboscene} formalizes HRI properties using tock-CSP and RoboScene, providing a UML-style notation for HRI scenarios. The conditions that they include in their models are likely instantiations of some of the categories of our taxonomy.



\new{A systematic literature review identifies 17 service robot design features 
under 3 headings: technical design, 
 HRI design, 
 and service system design 
 \cite{yang2026service}. 
 Several of their feature categories are mirrored in our taxonomy, including \category{Failure Recovery}  (Failure Impact Feature and Recovery Mechanism Feature in \cite{yang2026service}) and \category{System Capabilities} (Functional Capability Feature in \cite{yang2026service}). Others are higher-level: 
 e.g., Trust and Emotional Feature in~\cite{yang2026service} has no direct analog, though our \category{Information Sharing and Awareness} supports trust calibration. 
 Related work on recommender systems in social robots could help to create trustworthy systems but we did not find concrete examples in our search~\cite{huang2026reimagining}. We also did not find case studies where ethical requirements were present but this is an important aspect to be considered in future work~\cite{nam2026not}.}

\new{Another systematic review defines a taxonomy of robot autonomy for HRI \cite{kim2024taxonomy}. They identify 6 distinct forms of autonomy, grouped into 3 categories: robot and human involvement at runtime, human involvement before runtime, and expressions of autonomy at runtime. Our taxonomy does not distinguish between autonomy types (out of scope for this work) but the requirements in our corpora likely span them.} 

\section{\new{Construction Corpus}  Methodology}
\label{sec:methodology}

\begin{table}[t]
    \centering
    \scalebox{0.835}{
    \begin{tabular}{p{6.17cm} p{2.154cm} p{1.089cm}}
    \hline
         \textbf{Venue} & \textbf{Type} & \textbf{Ref.}\\ \hline\hline
         Advances in Industrial and Manufacturing Engineering & Journal  &\cite{segura2021human}\\\hline
         Conference on Human Factors in Computing Systems & Conference & \cite{agrawal2020next}\\\hline
         Towards Autonomous Robotic Systems & Conference & \cite{grigore2011towards}\\\hline
         IEEE Intelligent Systems & Journal & \cite{klien2005ten}\\\hline
         IEEE/AIAA Digital Avionics Systems Conference& Conference  & \cite{tokadli2022characteristics}\\\hline
        International Journal of Robotics Research & Journal  & \cite{webster2020corroborative}\\\hline
         IEEE Access & Journal  & \cite{damacharla2018common}\\\hline
         MITRE & Tech Report & \cite{mcdermott2018human}\\\hline
         International Requirements Engineering Conference & Conference  & \cite{jost2024digital}\\\hline
         ISO & Technical Standard &\cite{ISO-10218,ISO-102182-2,ISO-15066}\\\hline
         Federal Aviation Administration (FAA) & Technical Standard & \cite{HF-STD-001}\\\hline
         NASA & Technical Standard & \cite{nasa-std-3001}\\\hline
         Code of Federal Regulations (CFR) & Technical Standard & \cite{14cfr}\\\hline
         National Transportation Systems Center & Tech Report & \cite{yeh2016human}\\\hline
    \end{tabular}}
    
    \caption{Sources included in the \new{construction} corpus}
    \label{tab:sources}
    
\end{table}


We follow four stages: 1) literature and standards review to identify requirement sources, 2) requirement extraction and corpus assembly, 3) iterative taxonomy development, and 4) validation through expert evaluation and utility demonstration.

\textbf{Search Strategy:} We employed a hybrid, multi-stage approach. Initial sources were identified through expert recommendations and targeted searches of organization websites (e.g., NASA, FAA, and ISO), then expanded via snowballing~\cite{10.1145/2601248.2601268} through references in relevant technical standards, reports and publications. We also searched IEEE Xplore, ACM Digital Library, and Springer using combinations of core terms, e.g., ``human-robot team'', ``human-robot collaboration'', ``human-robot interaction'', combined with ``requirement'', ``specification'', ``standard'', ``safety''. We did not apply date restrictions, as foundational standards remain relevant. 

\textbf{Inclusion Criteria: }Sources were included if they: 
1) contained explicit requirements, design guidance or specification patterns; 2) addressed HRI in collaborative or team-based settings; 3) focused on operational, safety, and/or teamwork aspects; 4) provided sufficient detail to extract requirements.

\textbf{Exclusion Criteria: }
Sources were excluded if they: 1) focused solely on algorithmic performance without human interaction context; 2) addressed single-agent autonomous systems without human-robot collaboration; 3) provided high-level conceptual frameworks without specific requirements; 4) focused solely on shared goals or joint intentions in classical multi-agent systems literature (e.g.,~\cite{jarvis2006teams}). 
\textbf{Source Selection Process: }We began with 72 initial candidate sources. Following title and abstract screening, 42 documents were retained for full-text review. Application of the inclusion and exclusion criteria resulted in a \new{construction} corpus of 14 sources used for requirement extraction in Table~\ref{tab:sources}. 



\textbf{Extraction Process:} Two researchers independently reviewed each source to identify requirement statements. The researchers compared their extractions and discussed discrepancies. For ambiguous requirement statements, a third researcher reviewed to determine if the requirement statement followed the inclusion/exclusion criteria. Initial agreement between the primary extractors was 87\%. 
Disagreements primarily concerned: 1) Boundary between teamwork and general HRI; 
2) whether statements were requirements versus contextual information. 
For extraction, we distinguish teamwork requirements from general HRI requirements based on interdependent collaboration. We retained only requirements that explicitly support coordinated roles, shared responsibility, or distributed decision-making toward a common objective. Disagreements were resolved through discussion and third-party review, resulting in the final  \new{construction} corpus of 361 requirements.

\textbf{Taxonomy Development:} \new{Once we assembled our construction corpus, we examined and classified each requirement. Two existing frameworks were directly relevant to our work: the McDermott et al. Human-Machine Teaming (HMT) framework~\cite{mcdermott2018human} and NASA's Human-System Standard~\cite{nasa-std-3001}. McDermott et al. organize HMT into 10 themes across 4 higher-level categories (Transparency, Augmenting Cognition, Coordination, Design Specifics), providing generic requirements spanning autonomous vehicles to cognitive assistants that can be tailored to specific systems. NASA's standard defines human-system interaction requirement types spanning crew health, performance, training, and hardware/software design. These were developed through extensive expert consultation and operational use in safety-critical systems.  We also considered broader academic frameworks of HRI and CPS teaming (e.g., levels-of-autonomy~\cite{sheridan1978human} and interaction-mode classifications~\cite{klien2005ten}), but these describe conceptual collaboration modes rather than concrete requirement types.} 

\new{We adopted NASA's Human-System Standard~\cite{nasa-std-3001} as our initial resource because its categories are organized around concrete interaction types (e.g., automation system status provision, mode change notification) and written as specific requirements consistent with regulated specification practice~\cite{ISO29148:2018}. By contrast, McDermott et al.~\cite{mcdermott2018human}  organize categories around high-level HMT properties (e.g., Observability, Calibrated Trust, Common Ground), and the requirements within them are intentionally generic. NASA's Standard thus more closely matched the granularity of our corpus requirements and the RE practices our taxonomy is meant to support.}

Preliminary analysis revealed that NASA's framework was not enough for complete coverage of our \new{construction} corpus, necessitating extension and refinement. \new{We followed iterative refinement toward two ending conditions: an objective condition (all objects are classifiable without a residual category) and a subjective condition (categories are concise, robust, comprehensive, and meaningful to domain experts)~\cite{nickerson2013method}.} 
Categories from \cite{nasa-std-3001} were 1) broadened when few requirements matched a narrow category, and 2) narrowed when  requirements fitting a category had clear differences between them. 

\textbf{Taxonomy Structure:} The taxonomy derived from our \new{construction} corpus defines a two-level hierarchy, with high-level categories capturing abstract teamwork concerns, while low-level categories represent more concrete and actionable requirement groupings. 
\new{The two-level hierarchical structure is grounded in classification theory~\cite{kwasnik1999role}, where  hierarchical organization is appropriate when concepts can be meaningfully distinguished at different levels of abstraction.}
Each high-level category may comprise many low-level categories, while every low-level category maps exactly to one high-level category. 

\section{Taxonomy: High-Level Categories }
\label{sec:highleveltaxonomy}

Our categorization provides a high-level classification of HRT requirements 
(Fig.~\ref{fig:taxonomyCategories}, Table~\ref{tab:categories}). 
Each high-level category reflects a functional concern within teamwork and differs in terms of \emph{what is being exchanged} (e.g., data, commands, or safety measures), the \emph{purpose of the exchange} (e.g., situational awareness, control, or safety assurance), and the \emph{type of interaction} (one-way, two-way, proactive, or reactive). 

\category{Information Sharing and Awareness} addresses state visibility without prescribing action; \category{Control and Coordination} concerns allocation and authority;  \category{Communication and Decision Support} addresses influence on choice; \category{Safety and Risk Mitigation} focuses on protection and containment; \category{Team Performance Monitoring} concerns assessment of agent state; and \category{Team Capabilities and Authority} defines foundational properties. The first 3 categories primarily focus on \emph{real-time teamwork and decision-making}, ensuring shared understanding and synchronized action, whereas \category{Safety and Risk Mitigation} and \category{Team Performance Monitoring} focus on \emph{ensuring safe, reliable, and effective collaboration}. In addition, some categories (e.g., \category{Information Sharing and Awareness}) 
are predominantly proactive, while others (e.g., \category{Safety and Risk Mitigation}) 
define primarily reactive interactions. These distinctions ensure analytical separability while acknowledging operational interdependence - especially between \category{Information Sharing and Awareness}, which provides state visibility without directing behavior, and \category{Communication and Decision Support}, which actively shapes action. 



\begin{table*}[t]
    \centering
    \scalebox{0.86}{
    \begin{tabular}{p{6cm} p{4.3cm} p{4.3cm} p{4.7cm}}
    \hline
         \textbf{Category} & \textbf{Focus} & \textbf{Interaction Type} & \textbf{Key Differentiator}\\ \hline\hline
         \category{Information Sharing and Awareness} & Providing system status and data & Primarily one-way (status updates) & Ensures situational awareness but does not shape actions\\ \hline
         \category{Control and Coordination} & Managing tasks and system behavior & Two-way negotiation (commands and permissions) & Ensures the human defines automation's role and actions\\ \hline
         \category{Communication and Decision Support} & Aiding decision making & Interactive (recommendations and alerts) & Provides guidance rather than just raw data\\ \hline
         \category{Safety and Risk Mitigation} & Preventing/responding to hazards & Proactive and reactive (safety features, failure responses) &Focuses on preventing harm and handling failures \\\hline
          \category{Team Performance Monitoring} & Assessing team effectiveness & Continuous assessment (monitoring behavior)& Tracks ongoing performance rather than providing static data\\\hline
          \category{Team Capabilities and Authority} & Defining system capabilities and ensuring human authority & Static system design and emergency control &Establishes fundamental system abilities and human control mechanisms \\\hline
    \end{tabular}}
    
    \caption{High-level categorization of requirements}
    \label{tab:categories}
    \vspace{-15pt}
\end{table*}

For each category we provide a description of the (1) rationale; (2) type of exchanged information; (3) type of interaction and (4) key differentiator(s) between it and the other categories. Each category is decomposed into several subcategories, 
we describe these with examples in \S \ref{sec:lowleveltaxonomy}.

\subsection{\category{Information Sharing and Awareness}}
\label{sec:informationSharingCategory}

\noindent  \textbf{Rationale:} This focuses on the availability and accessibility of critical data for human and robotic teammates. It emphasizes maintaining situational awareness and ensuring that necessary data is provided to both parties for effective collaboration. \\
\textbf{Type of information:} Raw data and timely status updates.\\
\textbf{Purpose:} Ensuring both human(s) and robot(s) have adequate data to understand the system’s state and operate effectively.\\
\textbf{Type of interaction: }Primarily one-way. For example, robot providing data to human, or vice versa.\\
\textbf{Key Differentiator:} This category is about passive awareness. It ensures that operators are informed but it does not consider direct decision-making or control actions.\\ 
\textbf{Example Scenario:} The robot provides telemetry and automation status 
to ensure that the human has the required data.

\subsection{\category{Control and Coordination}}
\label{sec:controlAndCoordinationCategory}

\noindent \textbf{Rationale:} This category encapsulates the ability of human operators to configure, control, and delineate responsibilities within the Human-Robot team. It ensures that the robot is used and performs in a way that aligns with human intent while maintaining safety and operational efficiency.\\
\textbf{Type of information:} Commands, permissions, and role delineation between human and robot.\\
\textbf{Purpose:} Enabling the human to define, adjust, and/or authorize the robot's automated/autonomous actions while ensuring that the system operates within the defined framework.\\
\textbf{Type of interaction:} Two-way negotiation where humans set automation parameters and robots operate within them.\\
\textbf{Key Differentiator:} Unlike \category{Information Sharing and Awareness}, this category is about actively shaping how the robot functions within the team.\\ 
\textbf{Example Scenario:} The human sets robot automation levels, defines responsibilities and authorizes restarts after failures.

\subsection{\category{Communication and Decision Support}}
\label{sec:decisionSupportCategory}

\noindent \textbf{Rationale: }Requirements in this category center on the \emph{provision and clarity} of information. This category highlights the role of communication in facilitating decision-making and understanding between human and robot teammates. It includes requirements that are related to decision aids, interface clarity and notifications to support effective HRT.\\
\textbf{Type of Information:} Includes processed information, recommendations, and alerts from the robot to the human.\\
\textbf{Purpose:} Helping the human to make better decisions faster while maintaining situational awareness and trust in the robot.\\
\textbf{Type of Interaction: }  Interactive guidance, i.e., robot provides recommendations that the human interprets and acts upon.\\
\textbf{Key Differentiator:} Unlike  \category{Information Sharing and Awareness}, which provides data for passive awareness, this  captures data provision that actively shapes human decision-making. This includes on-demand decision aids and proactively volunteered data whose nature compels or directs human responses, differing from data updating situational awareness.\\
\textbf{Example Scenario:} A decision aid provides suggested actions, explains its reasoning, and notifies the human when it lacks sufficient data to provide recommendations.


\subsection{\category{Safety and Risk Mitigation}}
\label{sec:safetyAndRiskMitigationCategory}

\noindent \textbf{Rationale:} This comprises requirements that ensure safe operation within Human-Robot collaborative environments. It includes preventative safety measures, emergency responses, and failure recovery to protect human operators and robots.\\
\textbf{Type of information:} Safety, failure, and environmental safeguards between human, robot, and workspace.\\
\textbf{Purpose:} Ensuring safe HRT, minimizing the  risks to both human and robotic agents.\\
\textbf{Type of interaction:} Proactive safety measures like workspace design. Reactive safety measures like failure recovery.\\
\textbf{Key Differentiator:} Unlike \category{Control and Coordination}, which focuses on who does what, this category ensures that no matter who is in control, safety is prioritized at all times.
\\ 
\textbf{Example Scenario:} If the robot fails, it enters a safe mode and/or provides alerts to the human. Safety zones ensure that humans and robots operate without interference.

\subsection{\category{Team Performance Monitoring}}
\label{sec:performanceMonitoringCategory}

\noindent \textbf{Rationale:} This category covers the continuous assessment of both human and robot performance to maintain situational awareness and ensure optimal teamwork. It includes monitoring human behavior as well as robotic system performance.\\
\textbf{Type of information:} Observations of human and robot behavior, performance metrics, and real-time monitoring data.\\
\textbf{Purpose:} Ensuring humans and robots  perform optimally by adjusting operations based on observed performance.\\
\textbf{Type of interaction:} Continuous assessment where the robot monitors human behavior and vice versa.\\
\textbf{Key Differentiator:} While  \category{Information Sharing and Awareness}  provides static data, this focuses on continuous tracking and evaluation of human and robot performance.\\ 
\textbf{Example:} The system detects operator fatigue or degraded robot performance and adjusts operations accordingly.

\subsection{\category{Team Capabilities and Authority}}
\label{sec:capabilitiesCategory}

\noindent \textbf{Rationale:} Defines fundamental capabilities of robotic and human teammates, ensuring that robots are designed with characteristics to support effective collaboration, while guaranteeing that humans maintain ultimate system authority.\\
\textbf{Type of information:} System design considerations, human-focused requirements, and override mechanisms.\\
\textbf{Purpose:} Establishing the fundamental capabilities of both human and robot teammates while keeping human authority.\\
\textbf{Type of interaction:} Static design with emergency control: robot has built-in capabilities, while humans retain override.\\
\textbf{Key Differentiator:} Unlike \category{Control and Coordination}, which manages real-time interactions, this defines the fundamental capabilities and limits of human and robot teammates. \category{Control and Coordination} governs the runtime allocation of authority, whereas this category establishes the structural constraints and override boundaries for such allocation.
\\ 
\textbf{Example Scenario:} The robot’s designed capabilities (e.g., sensor accuracy, movement speed) determine how well it can collaborate. Meanwhile, the human retains the ability to override or shut down the robot at any time.\medskip


\section{Taxonomy: Low-level categories}
\label{sec:lowleveltaxonomy}

\begin{figure*}[t]
    \centering
    \includegraphics[width=0.88\linewidth]{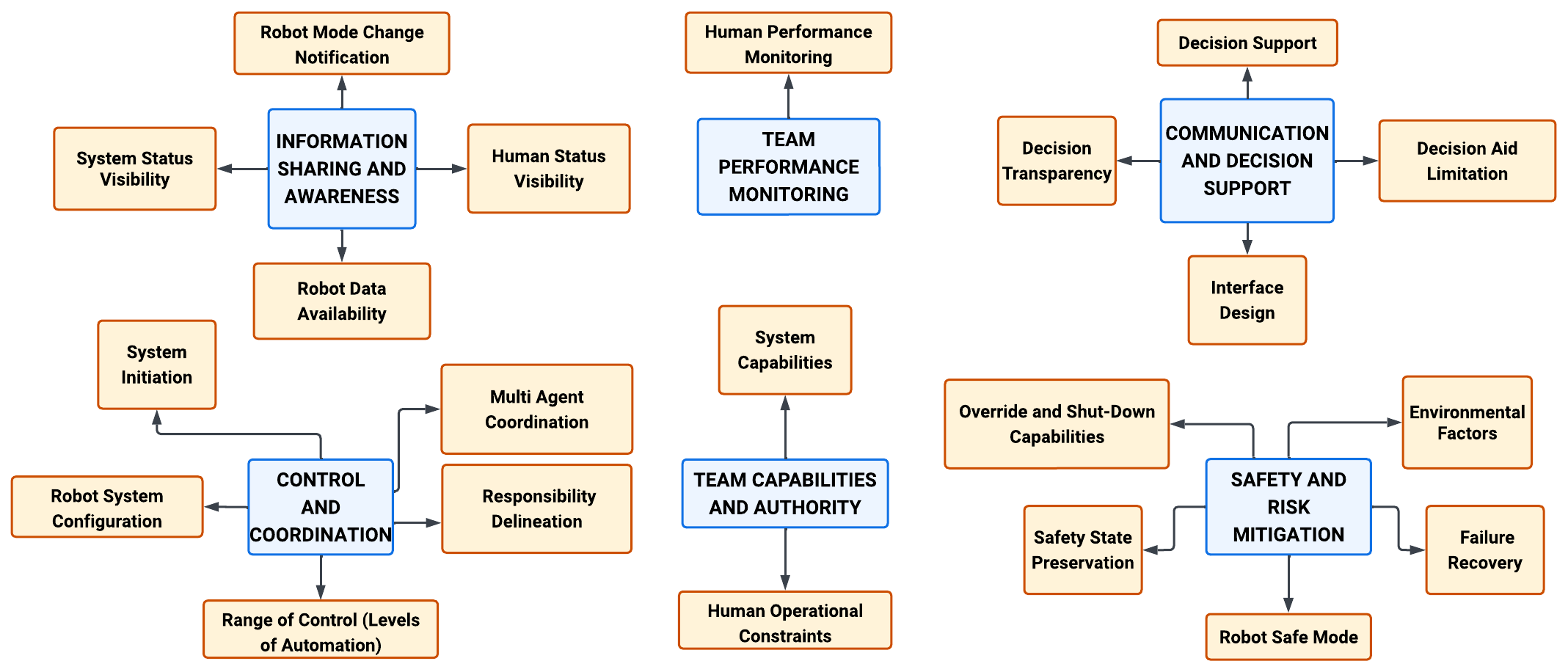}
    \caption{Hierarchy of categories in our taxonomy. The high-level categories are shown in the blue boxes and the low-level categories are in the orange boxes.}
    \label{fig:taxonomyCategories}
\end{figure*}

\noindent Fig. \ref{fig:taxonomyCategories} decomposes each high-level category into subcategories. 

\subsection{\category{Information Sharing and Awareness}}

\noindent \textit{1) Robot Data Availability}

\noindent \textbf{Rationale:} Data, such as telemetry, states, modes, etc., should be saved and made available to operators. 

\noindent \textbf{Example Requirement:}

\noindent \fbox{\parbox{0.48\textwidth}{ \small\textit{``Automated/robotic systems shall record and make available operational and performance data to both crew and ground support personnel.''} \cite{nasa-std-3001}}}\\

\noindent \textit{2) System Status Visibility}

\noindent \textbf{Rationale:} Human must maintain situational awareness to calibrate system trust and avoid errors. They need visibility of system health and projected state to assess robot performance. 

\noindent \textbf{Example Requirement:}

\noindent\fbox{\parbox{0.48\textwidth}{
\small\textit{``Automated/robotic systems shall provide the human operator with the following information: system state and projection of future state, including failure or decrements in performance (e.g., battery power versus traverse distance and assessment of uncertainty in projection of future state).''} \cite{nasa-std-3001}}}\\

\noindent \textit{3) Robot Mode Change Notification}

\noindent \textbf{Rationale:} Clear indication of the current mode helps prevent errors, where operators take inappropriate actions or fail to act due to misunderstanding the system's state. 
Appropriate notification allows the operator to prepare for a mode change.

\noindent \textbf{Example Requirement:}

\noindent\fbox{\parbox{0.48\textwidth}{\small\textit{``The automated/robotic system shall notify the human operator of mode changes of any safety-critical operations.''} \cite{nasa-std-3001}}}\\

\noindent \textit{4) Human Status Visibility}

\noindent \textbf{Rationale:} As well as passing information from robot to humans (e.g., \textit{System Status Visibility}), it is crucial that there is a way of transferring information from human to robot. This subcategory encapsulates the requirements relating to robots that must interpret and receive information from humans.

\noindent \textbf{Example Requirement:}

\noindent\fbox{\parbox{0.48\textwidth}{\small\textit{``The automation/autonomy shall enable users to input informal priorities and changes in those priorities.''} \cite{mcdermott2018human}}}

\subsection{\category{Control and Coordination}}

\noindent \textit{1) System Initiation}

\noindent \textbf{Rationale:} The human operator is to remain in control and must authorize a system restart after an emergency stop.

\noindent \textbf{Example Requirement:}

\noindent \fbox{\parbox{0.48\textwidth}{ \small\textit{``If a detected failure occurs, autonomous operation resumes after a deliberate restart from outside the collaborative workspace.''} \cite{ISO-102182-2}}}\\

\noindent \textit{2) Robot System Configuration}

\noindent \textbf{Rationale:} The human operator must retain control and be able to modify automation configuration parameters, such as setup inputs, initial conditions, and termination conditions. 
However, certain configurations must remain restricted due to system-specific performance or safety constraints. 

\noindent \textbf{Example Requirement:}

\noindent \fbox{\parbox{0.48\textwidth}{ \small\textit{``Automated/robotic systems shall provide the human operator the ability to modify system configuration within the safety and performance limits of the system.''} \cite{nasa-std-3001}}}\\

\noindent \textit{3) Responsibility Delineation}

\noindent \textbf{Rationale:} A clear  delineation of the responsibilities of the human and robot is vital to ensure that the operator performs appropriate and timely actions. Transitions between robotic and human control must execute under well-defined procedures that specify the system state before and after  transitions. 

\noindent \textbf{Example Requirement:}

\noindent \fbox{\parbox{0.48\textwidth}{ \small\textit{``Automated/robotic systems shall indicate whether a human operator or system is expected to perform a particular operation at a specific time.''} \cite{HF-STD-001}}}\\

\noindent \textit{4) Multi-Agent Coordination} 

\noindent \textbf{Rationale:} Requirements specifically relating to the interaction between more than one agent in a team, e.g., 
the connection between and sharing of information amongst systems. This includes multi-robot and multi-human coordination.

\noindent \textbf{Example Requirement:}

\noindent \fbox{\parbox{0.48\textwidth}{\small \textit{``The automated/robotic system shall make connections between requests from different agents, both human and automated.''} \cite{mcdermott2018human}}}\\

\noindent \textit{5)  Range of Control (Levels of Automation) }

\noindent\textbf{Rationale:} Robotic systems are expected to be operated in different manners; e.g., controlled directly (i.e., manual control) or commanded remotely (i.e., supervisory control). The multiple options associated with these systems must be operable via the range of controls available to the human operator.

\noindent \textbf{Example Requirement:}

\noindent \fbox{\parbox{0.48\textwidth}{ \small\textit{``Adaptive automation should be implemented at the point at which the user ignores a critical amount of information.''} \cite{HF-STD-001}}}\\


\subsection{\category{Communication and Decision Support}}

\noindent \textit{1)  Decision Support}

\noindent \textbf{Rationale:} Decision aids provide pertinent information, analysis, and/or suggested solutions for continued operations. The system ultimately needs to enable the operator to make those decisions, whether or not it is the operator that acts on them.

\noindent \textbf{Example Requirement:}

\noindent \fbox{\parbox{0.48\textwidth}{ \small\textit{``The automated/robotic system shall provide enough information about the developing event so that the operator can intervene effectively.''} \cite{mcdermott2018human}}}\\


\noindent \textit{2) Decision Transparency} 

\noindent \textbf{Rationale:} The human must understand why the robot recommends actions, and the action consequences, to make informed decisions without significantly disrupting the operator's task, maintaining situational awareness, and calibrating trust.

\noindent \textbf{Example Requirement:}

\noindent \fbox{\parbox{0.48\textwidth}{\small\textit{``Decision aid systems shall provide explanations and rationales, and consequences of potential actions.''} \cite{nasa-std-3001}}}\\


\noindent \textit{3) Decision Aid Limitation}

\noindent \textbf{Rationale:} Human operators must be alerted when a decision aid cannot assist due to lack of data or design limitations so that data can be provided, or other avenues explored promptly.

\noindent \textbf{Example Requirement:}


\noindent \fbox{\parbox{0.48\textwidth}{ \small\textit{``Decision aids shall notify the human operator when a problem or situation is beyond the aid’s capability.''} \cite{nasa-std-3001}}}\\

\noindent \textit{4) Interface Design}

\noindent \textbf{Rationale:} Communication through displays and interfaces, including how and when warnings should be presented.

\noindent \textbf{Example Requirement:}

\noindent \fbox{\parbox{0.48\textwidth}{ \small\textit{``Warning and caution alerts must provide timely attention-getting cues through at least two different senses by a combination of aural, visual, or tactile indications.''} \cite{14cfr}}}\\

\subsection{\category{Safety and Risk Mitigation}}

\noindent \textit{1) Environmental Factors}

\noindent \textbf{Rationale:} Collaborative workspaces must prioritize operator safety and efficiency. Environment considerations such as clearly defined collaborative zones, perimeter safeguarding, and protective measures ensure that robots operate safely within shared spaces, allowing for controlled entry and human oversight. These safeguards create a structured and adaptive environment where humans and robots work together securely.

\noindent \textbf{Example Requirement:}

\noindent \fbox{\parbox{0.48\textwidth}{ \small\textit{``The design of the collaborative workspace shall be such that the operator can easily perform all tasks and the location of equipment and machinery shall not introduce additional hazards.
''} \cite{ISO-10218}}}\\

\noindent \textit{2) Safety State Preservation}

\noindent \textbf{Rationale:} In a failed transfer of control from the robot to the human, the robot experiencing degraded performance must take measures to remain in a safe state, including alerting the operator of the degraded state and transfer of control. 

\noindent \textbf{Example Requirement:}

\noindent \fbox{\parbox{0.48\textwidth}{ \small\textit{``Automated/robotic systems not selected for operation remain in a safe state in a multi-robot cell.''} \cite{ISO-102182-2}}}\\


\noindent \textit{3) Robot Safe Mode}

\noindent \textbf{Rationale:} Protective actions include avoidance maneuvers, and protective stops if the human operator can no longer command the system. The transition to operator control needs to occur safely and minimize harm to the operator and vehicle.

\noindent \textbf{Example Requirement:}

\noindent \fbox{\parbox{0.48\textwidth}{ \small\textit{``The automated/robotic system shall take protective action (e.g., avoidance maneuver, protective stop) or request that the operator safely take control if the system's operational safety threshold is exceeded.''} \cite{nasa-std-3001}}}\\

\noindent \textit{4) Failure Recovery}

\noindent \textbf{Rationale:} If a failure  occurs, the data provided by the system must enable the human operator to collect confirming/exclusion evidence to determine a safe course of action.

\noindent \textbf{Example Requirement:}

\noindent \fbox{\parbox{0.48\textwidth}{ \small\textit{``Early warning notification of pending automation failure or performance decrements should use estimates of time needed for the user to adjust to task load changes due to automation failure.''} \cite{HF-STD-001}}}\\

\noindent \textit{5) Override and Shut-Down Capabilities}

\noindent \textbf{Rationale:} The system allows the human to override/shut down robotic systems if  they present a risk, or if redirection is needed. 
It is vital that the override or shut down is performed safely, i.e., avoids inadvertent harm to crew and vehicle.

\noindent \textbf{Example Requirement:}

\noindent \fbox{\parbox{0.48\textwidth}{ \small\textit{``The automated/robotic system shall provide the operator the capability to override the automation and assume partial or full manual control of the system to achieve operational goal states.''} \cite{mcdermott2018human}}}\\

\subsection{\category{Team Performance Monitoring}}

\noindent \textit{1) Human Performance Monitoring}

\noindent \textbf{Rationale:} The robotic agent in the team must monitor aspects of human behavior, performance and abilities to maintain understanding of the situation/tasks. \new{The robot must monitor the situational awareness of its human teammate(s)  so that they can take actions to improve this awareness if needed \cite{ali2025estimating}.}

\noindent \textbf{Example Requirement:}

\noindent \fbox{\parbox{0.48\textwidth}{ \small\textit{``The autonomous teammate shall monitor the actions and performance of the human teammate(s).''} \cite{tokadli2022characteristics}}}\\

\subsection{\category{Team Capabilities and Authority}}

\noindent \textit{1) System Capabilities}

\noindent \textbf{Rationale:} Robot capabilities and characteristics are critical in HRT to determine the system’s ability to operate reliably, predictably, and collaboratively with human teammates. These requirements represent capabilities and characteristics that are essential for collaboration in teams.

\noindent \textbf{Example Requirement:}

\noindent \fbox{\parbox{0.48\textwidth}{ \small\textit{``Automation should be designed to adapt by providing the most help during times of highest user workload, and somewhat less help during times of lowest workload.''} \cite{HF-STD-001}}}\\


\noindent \textit{2) Human Operational Constraints}

\noindent \textbf{Rationale:} Requirements that focus on the human counterpart rather than the robotic system. This includes requirements relating to the actions and behaviors that a human should withhold, and system requirements that aid the human teammate.

\noindent \textbf{Example Requirement:}


\noindent \fbox{\parbox{0.48\textwidth}{ \small\textit{``Users should be given an adequate understanding of how the automated system works in order to monitor effectively.''} \cite{HF-STD-001}}}\\

\section{Metrics and Expert Evaluation}
\label{sec:expertevaluation}
The full list of collected requirements, the slides \new{and interview protocol} used for expert interviews are available in~\cite{teamworktaxonomy2024}.

\subsection{Category and Subcategory Metrics}

\label{sec:metrics}
\begin{table}[t]
\centering
    \scalebox{0.85}{
    \begin{tabular}[t]{p{5.2cm} |p{0.6cm}| p{0.65cm} p{0.5cm} p{0.65cm}}
    \hline
         \textbf{Category} & \textbf{Total} & \textbf{Human-Led} & \textbf{Robot-Led} & \textbf{Shared}\\ \hline\hline
         \textsc{Information Sharing and Awareness} & \textbf{56} & \textbf{N} & \textbf{Y} & \textbf{Y} \\ \hline
        \rowcolor{lightgray!60} \textit{Robot Data Availability} &6 & \textbf{-} & \textbf{Y} & \textbf{-}\\
         \hline
      \rowcolor{lightgray!60}  \textit{System Status Visibility} &35 & \textbf{-} & \textbf{Y} & \textbf{-}\\ \hline
         \rowcolor{lightgray!60}  \textit{Robot Mode Change Notification} &10 & \textbf{-} & \textbf{Y} & \textbf{-}\\ \hline
          \textit{Human Status Visibility} &5 & \textbf{Y} & \textbf{-} & \textbf{Y} \\ \hline \hline
          \textsc{Control and Coordination} &\textbf{59} & \textbf{Y} & \textbf{N} & \textbf{Y} \\ \hline
          \rowcolor{lightgray!60}  \textit{System Initiation} &4 & \textbf{Y}  & \textbf{-} & \textbf{-}\\ \hline
          \rowcolor{lightgray!60}  \textit{Robot System Configuration} &14 & \textbf{Y} & \textbf{-} & \textbf{-}\\ \hline
          \rowcolor{lightgray!60}  \textit{Responsibility Delineation} &13 & \textbf{Y} & \textbf{Y} & \textbf{Y} \\ \hline
          \textit{Multi-Agent Coordination} & 5 &\textbf{Y} & \textbf{Y}& \textbf{Y}\\ \hline 
          \rowcolor{lightgray!60}  \textit{Range of Control (Levels of Automation)} &23 & \textbf{Y} & \textbf{-} & \textbf{-} \\ \hline\hline
  \textsc{Communicat{\-}ion and Decision Support} &\textbf{100}& \textbf{Y} & \textbf{Y} & \textbf{Y}\\ \hline
          \rowcolor{lightgray!60}  \textit{Decision Support} &49 & \textbf{Y} & \textbf{Y} & \textbf{Y} \\ \hline
          \rowcolor{lightgray!60}  \textit{Decision Transparency} &23 &\textbf{-} & \textbf{Y} &\textbf{-} \\ \hline
          \rowcolor{lightgray!60}  \textit{Decision Aid Limitation} &3 &\textbf{-} & \textbf{Y} & \textbf{-} \\ \hline
          \textit{Interface Design} &25 &\textbf{Y} &\textbf{Y} &\textbf{Y} \\ \hline \hline
          \textsc{Safety and Risk Mitigation} &\textbf{61} & \textbf{Y} & \textbf{Y} & \textbf{Y} \\ \hline
          \textit{Environmental Factors} &11&\textbf{Y} & \textbf{Y} & \textbf{Y} \\ \hline
          \rowcolor{lightgray!60}  \textit{Safety State Preservation} &10 & \textbf{Y} & \textbf{Y} & \textbf{Y} \\ \hline
          \rowcolor{lightgray!60}  \textit{Robot Safe Mode} &18 & \textbf{-} & \textbf{Y} & \textbf{-}\\ \hline
          \rowcolor{lightgray!60} \textit{Failure Recovery} & 11&\textbf{Y} & \textbf{Y} & \textbf{Y} \\ \hline
          \rowcolor{lightgray!60}  \textit{Override and Shut-Down Capabilities} &11& \textbf{Y} & \textbf{-} & \textbf{Y} \\ \hline \hline
          \textsc{Team Performance Monitoring} &\textbf{9} & \textbf{-} & \textbf{Y}  & \textbf{-}\\ \hline
          \textit{Human Performance Monitoring} &9& \textbf{-} & \textbf{Y} & \textbf{-}\\ \hline \hline
        \textsc{Team Capabilities and Authority} &\textbf{76}& \textbf{Y} & \textbf{Y} & \textbf{Y} \\ \hline
        \textit{System Capabilities} &64& \textbf{Y} & \textbf{Y} & \textbf{-}\\ \hline
          \textit{Human Operational Constraints} &12 & \textbf{Y} & \textbf{-} & \textbf{-}\\ \hline \hline

    \end{tabular}}

    \caption{The category distribution of  our  \new{construction} corpus. We classify categories as Human-Led, Robot-Led or Shared. The total requirements for each  category is the sum of its subcategories. Gray shading indicates an analog in 
    \cite{nasa-std-3001}.} 
\label{tab:numberpercategory}
    \label{tab:humanledVsRobotledCategories}
\end{table}

Table \ref{tab:numberpercategory} collates the 361 requirements in our \new{construction} corpus on a per taxonomy category (\S \ref{sec:highleveltaxonomy}) and subcategory (\S \ref{sec:lowleveltaxonomy}) basis. 
Most requirements (100/361) fall into \category{Communication and Decision Support}, mainly in the \textit{Decision Support} subcategory. 
The next largest category was \category{Team Capabilities and Authority}, primarily comprising the \textit{System Capabilities} subcategory. Many of these addressed handover tasks between humans and robots, e.g.,~\cite{webster2020corroborative}. 

Several requirements in \category{Safety and Risk Mitigation} concerned safety mechanisms for robots operating in close proximity to humans, including provisions specific to handover tasks. 
In \textit{Environmental Factors}, we found several requirements that referred to the design of the collaborative workspace and physical protective measures. These requirements are also relevant in multi-robot teaming situations.

Within \category{Control and Coordination}, many requirements fell into the \textit{Range of Control (Levels of Automation)} subcategory. Many of these were concerned with the robot moving through different degrees of automation/autonomous functionality ranging from tele-operated to fully autonomous. 

The largest subcategory within \category{Information Sharing and Awareness} was \textit{System Status Visibility}. This is unsurprising as the human must remain aware of the robot's current state, to make informed decisions using their knowledge and trust in the robot. 
Several requirements specified that the robot maintain awareness of the human teammate’s status, which we classified as \textit{Human Status Visibility}. We found 9 requirements on \textit{Human Performance Monitoring}. We did not find any on  monitoring robot performance \new{in the construction corpus.} 


Table~\ref{tab:numberpercategory} indicates (shaded rows) taxonomy categories that have an analogous category in~\cite{nasa-std-3001}, which contributed 14 requirement categories and multiple requirement examples to our taxonomy. We critically refined several of these, updating terminology based on evidence from our search and expert evaluation. For example, we replaced \textit{Automation System Status Provision}~\cite{nasa-std-3001} with the \textit{System Status Visibility} subcategory. We replaced `Automation' with `Robot' across several categories to better reflect the scope of requirements that we identified, which extend beyond automation including broader system components. We adopt `Visibility' in place of `Provision' to more accurately capture the intent of these requirements and to address terminology that consistently caused confusion during expert review. Our taxonomy extends beyond \cite{nasa-std-3001} by introducing 7 new categories derived from the wider literature. These categories have no counterpart in existing standards and represent a substantive original contribution to the systematic characterization of HRT requirements.

\subsection{Expert Evaluation}
\label{sec:eval}
To validate the taxonomy, we interviewed 5 experts with backgrounds in HRT, Human Factors and RE (2 academics, 2 from large robotics engineering companies and 1 from a space agency).  
We examined 
conceptual validity (\textbf{RQ1}) and practical applicability (\textbf{RQ2}). 
Participants were guided through the 6 high-level categories and asked 
about conceptual clarity and distinctness, subcategory classification rationale, and whether each category captured the intended class of HRT requirements (\textbf{RQ1}). For \textbf{RQ2}, participants assessed coverage and potential gaps by relating categories to real HRT systems and reflected on the taxonomy's usefulness for requirements elicitation and specification analysis. We thematically analyzed~\cite{lochmiller2021conducting} the discussions into 5 topics: 


\noindent \textbf{1) Awareness vs Action:} Experts recognized the separation between awareness and action as a useful operational distinction (\category{Information Sharing and Awareness}, and \category{Communication and Decision Support}, respectively). 
However, they noted that information initially intended to support passive understanding can, depending on the context, become relevant or urgent to decisions. As such, systems may need mechanisms to signal changing criticality while still preserving the conceptual boundary between awareness/informing and acting/directing. 

Several experts emphasized the importance of shared Situational Awareness (SA) \cite{endsley2017toward} relating to team members' (human and robot) internal models of external world \cite{endsley1988design}. SA helps to ensure that both humans and robots maintain a shared and accurate understanding of the environment, tasks, and each other's states \cite{nasa-std-3001}. This enables safe coordination, decision-making, and effective responses to unexpected changes. \new{SA was not included as a stand-alone taxonomy category as it is a cross-cutting cognitive outcome rather than a discrete functional concern. Confining it to a single category would obscure the distinct requirement types that collectively support it across the taxonomy.} We decompose SA into operational requirement classes \cite{endsley1995taxonomy}. \category{Information Sharing and Awareness} supports level 1 (perception) through visibility of system state, \category{Communication and Decision Support} supports level 2 (comprehension) by providing structured rationale and interpretations, and \category{Team Performance Monitoring} supports aspects of level 3 (projection) by anticipating performance decrements. However, projection in Human-Robot teams extends beyond these decrements and requires \textit{anticipation} of authority transition and shifts in responsibility and control, which is not fully represented in our taxonomy. Requirements that enable structured degradation and explicit authority delineation therefore operationalize Level 3 SA as a coordination property rather than a cognitive one, suggesting that SA in Human-Robot teams is closely connected with authority management. 



\noindent \textbf{2) Reduced Human to Robot Information and Negotiated Control:} While the taxonomy and extracted requirements provide structure for information flow from robot to human, there is less detail of what data the robot should have about its human teammate. Experts highlighted the importance of representing human priorities, condition, and behavior to enable adaptation and coordination. Further, they challenged the implicit assumption that control flows primarily from human to robot, noting situations where the robot must guide, influence, or temporarily supersede human action, especially during error, uncertainty, or time pressure. These observations reframe control as a dynamic, bidirectional exchange grounded in reciprocal awareness rather than a fixed allocation of authority. 
 

\noindent \textbf{3) Ergonomics and Human Variability:} Safety cannot be reduced to collision avoidance or fail-safe mechanisms; workspace design, human variability, and fatigue considerations must also be incorporated within ergonomic requirements. Experts further emphasized that effective teamwork requirements must accommodate variability in human behavior and attention, rather than relying on assumptions of consistent compliance. Together, these observations highlight the need to account for physical, cognitive, and behavioral variability in the design of safe and effective human-robot collaboration.


\noindent \textbf{4) Human Performance Monitoring Challenges:} 
All experts regarded this category as essential but challenging. Measuring physical performance is feasible but assessing cognitive state, intention, or understanding is more difficult. One expert extended this further by suggesting future systems may need explainability of human behavior. 

\noindent \textbf{5) Additional Categories:} Within \category{Team Performance Monitoring}, several experts suggested further refinement into subcategories such as \textit{Human Performance}, \textit{Robot Performance}, \textit{Quality of Interaction}, and \textit{Monitoring of Operational Parameters}. These suggestions reflect a desire to distinguish between physical performance metrics, cognitive or behavioral assessment, and interaction-level dynamics. 
\new{The utility demonstration in \S\ref{sec:utilitydemonstration} provides empirical support for the experts' intuition, with requirements involving robot monitoring identified as falling outside the existing subcategory, corroborating it as a candidate for future extension.} 

The experts viewed the taxonomy as conceptually coherent and practical for requirement elicitation. They also identified areas for refinement. 
\new{As mentioned in \S\ref{sec:metrics}, we made modifications to several taxonomy categories in light of the feedback from the expert reviewers. We detail the changes in~\cite{teamworktaxonomy2024}.
}

\subsection{Human-Led, Robot-Led and Shared Classification}


To answer \textbf{RQ2}, we analyze the classification from three operational perspectives: (1) \textbf{Human-Led}, (2) \textbf{Robot-Led} and (3) \textbf{Shared} (Table~\ref{tab:humanledVsRobotledCategories}) to expose patterns of authority and interdependence often implicit in existing standards. 
This framing clarifies how authority boundaries shape safe collaboration.

\subsubsection{Structural Boundaries of Authority}

Our ternary classification aligns with established models of autonomy such as Sheridan's 10 Levels of Automation (LoA)~\cite{sheridan1978human}. We map Human-Led to Low LoA (1--4), Robot-Led to High LoA (7--10), and Shared to Mid LoA (5--6). This operational analysis  reinforces the distinction between runtime authority allocation (\category{Control and Coordination}) and structural authority constraints (\category{Team Capabilities and Authority}).

Failure handling exemplifies the boundary. Detection is Robot-Led (the robot autonomously enters \emph{safe mode} or executes \emph{failure recovery}), while choice of response is typically Human-Led, and execution is Shared. For effective handover, \emph{Failure Recovery} requires the robot to convey error severity and type rather than generic failures, so that the human can classify the failure progression and respond accordingly.


Human-Led subcategories include the authority to override or shutdown the robot in unsafe situations (\emph{Override and Shut-Down Capabilities}) which is essential to meet legal and ethical requirements, especially in regulated domains. Most subcategories within \category{Control and Coordination} are Human-Led, reflecting the need for retained human supervisory authority. This override authority interacts with the Robot-Led safety functions in \category{Safety and Risk Mitigation} (e.g., \emph{Robot Safe Mode}). While the robot can autonomously stabilize or enter a safe state, human override remains the final control layer when automated safeguards fail. 
This supports work on shared control in HRI that emphasizes dynamic arbitration of control authority through intent detection and feedback  \cite{losey2018review}. These authority boundaries are especially critical during failure, when control  transitions under time and safety constraints.

\subsubsection{Graceful Degradation} 
A critical concern in Robot-Led teamwork is graceful degradation \cite{ishigooka2019graceful}, i.e., the ability of the system to fail safely and support a structured, predictable transition of control to the human under failure or performance decrement. As robot autonomy increases, the human operator becomes progressively removed from active control, making transition quality increasingly consequential. 
Requirements addressing graceful degradation are captured primarily within \category{Safety and Risk Mitigation}. 
%
Effective safety degradation requires explicit specification of how authority transitions during failure, who stabilizes, who diagnoses, and who determines subsequent action.  
Shared safety protocols and structured responsibility shifts ensure that control transfer occurs predictably and without introducing additional risk, whether that be during Human-Led, Robot-Led or Shared tasks.


\subsubsection{Shared Awareness}
Beyond authority transitions, effective teamwork requires reciprocated modeling between agents. The lack of requirements concerning human cognitive state, adaptability and reciprocal awareness suggests that existing standards prioritize system functionality over modeling the human as an active teammate.
For example, if the robot detects elevated cognitive strain in the human, (within \emph{Human Performance Monitoring}), it may adjust its communication frequency or physical behavior accordingly. 
Expert feedback highlighted both the limited representation of human cognitive modeling in current requirements and the difficulty of specifying observable indicators of intention, understanding, or trust.
Recent work \cite{deters2025identifying} explores user indicators that may prompt explanations, suggesting a pathway toward specifying observable measures of mutual understanding.

A related challenge is vigilance decrement~\cite{greenlee2019driver,hancock2026human}, i.e., 
prolonged low workload or monotonous activity reduces alertness and increases susceptibility to error. Requirements for human performance monitoring must capture both overload and under-stimulation.  \category{Communication and Decision Support} is Shared, 
and its effectiveness depends on the human's ability to understand the rationale behind robot recommendations. 

\subsubsection{Interdependence}
Shared functions regulate interdependence by coordinating communication, role delineation and risk management, to support efficient and effective teamwork. These requirements manage the complexities of interdependence, demanding that both (human and robot) agents engage in mutual communication and coordination to mitigate shared risk and ensure role compatibility. For example, \category{Control and Coordination} is shared when role delineation and transition procedures are in place, and \category{Information Sharing and Awareness} is shared when status updates are bidirectional. 
When functions are shared between human and robotic agents, clear delineation of authority is essential. Ambiguity over who is ultimately in charge during shared tasks contributes to coordination failures and safety incidents \cite{fuchsoptimizing2024}. 


Many of the categories and subcategories are \textit{Shared} (Table \ref{tab:humanledVsRobotledCategories}), which underscores the need for a comprehensive delineation of responsibilities in Human-Robot teamwork scenarios. This undoubtedly extends to teams that involve multiple human and robot agents working to achieve a shared goal. These would be explicitly considered in the \textit{Multi-Agent Coordination} and  \textit{Responsibility Delineation} categories. 


These observations suggest teamwork depends on authority structures, fault management, and mutual coordination rather than on hardware setup or the number or type of agents.


\section{\new{Utility Demonstration}}
\label{sec:utilitydemonstration}



\new{We conducted a utility demonstration, per Usman et al.'s guidance on taxonomy validation~\cite{usman2017taxonomies},  as a second measure to validate our taxonomy.}

\subsection{\new{Construction of an Independent Validation Corpus}}

\new{We assembled a second corpus (available in~\cite{teamworktaxonomy2024}) of requirements  from 19 sources, 
spanning 6 HRT domains: Aerospace \& Aviation~\cite{fuchs2025safe, westin2025human}, Emergency \& First Response~\cite{streiffertautonomous, pratt2006conops, humphrey2015human, hagenow2024system}, Healthcare~\cite{wang2026bidirectional, kubota2022cognitively, taylor2025rapidly}, Industrial Manufacturing~\cite{gualtieri2024updating, gualtieri2020safety, wu2025risk, boonyard2026speaking, yang2025implicit}, Defense \& Security~\cite{shively2017human, ye2024autonomy}, and Social \& Service Robotics~\cite{stange2022self, kim2026speaking,xu2026designing}. These were not used to construct the taxonomy in \S III. Sources were selected via purposive sampling~\cite{etikan2016comparison}, to evaluate  applicability across diverse  contexts rather than achieving exhaustive coverage.} 

\new{Source identification used two strategies. First, we conducted targeted searches of Google Scholar and NASA's Technical Reports Server (NTRS) using the search terms (individually and combined): \textit{CONOPS, requirements, technical document, rover, robot, ARTEMIS requirements, human-machine interaction}. Second, we manually reviewed proceedings from the International Conference on Human-Robot Interaction (HRI) (2016--2026), extracting requirement statements from 9 relevant papers. 
We chose HRI as a top peer-reviewed venue to ensure the validation corpus reflects representative human-robot interaction research.
}
\new{To ensure that any inability to classify a requirement could be attributed to gaps in the taxonomy rather than methodological drift between the corpora, we reused the inclusion and exclusion criteria from \S\ref{sec:methodology}.}

\subsection{\new{Extraction and Classification Procedure}}

\new{We followed the same extraction protocol used for the construction corpus in \S\ref{sec:methodology}. Two researchers independently extracted candidate requirements and applied the inclusion criteria, with an initial inclusion agreement of 91.09\%, 
compared to 87\% for the construction corpus. 
}
\new{From an initial 823 candidate statements, 448 were retained as in-scope after filtering, 375 were excluded as they described single-agent autonomy or contexts different than teamwork requirements.}

\new{Requirements were independently categorized by two researchers, achieving 94\% initial agreement. 
Disagreements clustered at boundaries between adjacent categories rather than being distributed randomly, indicating that classification was straightforward for most requirements and ambiguity arose where two categories had overlapping concerns. These were resolved through discussion against the category definitions.}

\subsection{\new{Coverage results}}
\new{Table~\ref{tab:utility-distribution} shows the distribution of 448 requirements across taxonomy subcategories and domains. Classification was achieved for 91.9\% (412/448) of in-scope requirements using the existing taxonomy, demonstrating that the 6 high-level categories and 21 subcategories provide broad coverage across domains not in the construction corpus. The remaining requirements could each be allocated a high-level category but did not fit an existing subcategory (beige rows, Table~\ref{tab:utility-distribution}) 
and are classed as candidates for taxonomy extension.}\smallskip

\noindent\new{\textbf{Convergence with construction corpus:} The distribution of requirements closely mirrors that of the construction corpus (Table~\ref{tab:numberpercategory}). \category{Communication and Decision Support} remains the largest category (29.2\% here and 27.7\% in the construction corpus), followed by \category{Team Capabilities and Authority} (23.2\% here and 21.1\% in the construction corpus). The \category{Team Performance Monitoring} is again the sparsest (4.7\% here and 2.5\% in the construction corpus). The three middle-ranked categories reorder slightly but each remains within 13--17\% in both corpora. The convergence of these patterns across two independently assembled corpora suggests that they reflect the structure of the field rather than artifacts of source selection.}\smallskip

\noindent\new{\textbf{Domain-specific characteristics:} Table~\ref{tab:utility-distribution} shows that each domain focuses on different categories. Industrial Manufacturing is led by \category{Safety and Risk Mitigation}, where most requirements (39/43) belong in the \textit{Environmental Factors} subcategory, reflecting shared workspace design standards~\cite{ISO-10218}. 
Healthcare is led by \category{Communication and Decision Support}, as these requirements (42\%) often define how data is shown to clinicians/patients. This is mirrored in Social \& Service Robotics with 38.7\% of requirements in this category. In contrast, Emergency \& First Response had a low incidence of requirements in this category (5\%) with \category{Team Capabilities and Authority} being its most populous category (49\%). Defense \& Security focuses on \category{Control and Coordination} reflecting the importance of adjustable range of automation in mission-critical settings. 
These patterns show that the taxonomy generalizes across domains without flattening their differences, supporting its use as a shared reference framework.}\smallskip

\noindent\new{\textbf{Sparsely populated categories:} Several subcategories remain sparsely populated in both corpora. \category{Team Performance Monitoring} contains only 21 requirements in the validation corpus, being completely absent from Social \& Service Robotics. 
This persistent under-representation, observed across both corpora and corroborated by expert evaluation (\S\ref{sec:eval}), suggests that human-side monitoring is under-specified in current practice. This illustrates how the taxonomy can act as a diagnostic instrument that surfaces systematic absences and directs elicitation effort toward the concerns most likely to be missing from a specification. A similar pattern holds for \textit{Failure Recovery} (1/448), \textit{System Initiation} (3/448), \textit{Override and Shutdown Capabilities} (4/448), and \textit{Decision Aid Limitation} (no representation in the validation corpus). These subcategories collectively describe how a Human-Robot team behaves at the boundary of nominal operation, meaning at failure, restart and handover. Their consistent under-representation across different sources indicates a genuine specification gap.}\smallskip

\noindent\new{\textbf{Candidates for taxonomy extension:} Some validation-corpus requirements (36/448) fit a high-level category but not any existing subcategory. 
These are candidates for extension rather than signs of a structural weakness since all requirements were classifiable at the high-level category. 
From preliminary analysis, these fall into candidate subcategories: 
requirements for assessing robot performance as a symmetric counterpart to \textit{Human Performance Monitoring} (n=6 in \category{Team Performance Monitoring}); requirements concerning the robot's social presentation and sustaining user engagement over time, rather than any specific information exchange or decision (n=21 in \category{Communication and Decision Support}); and requirements relating to explainability and human acceptance that sit below the interaction level addressed by \category{Communication and Decision Support}, concerning instead structural prerequisites for explainability and conditions for accepting robotic teammates (n=9 in \category{Team Capabilities and Authority}). Full definitions and empirical validation of these subcategories is future work. }

\begin{table}[t]
\centering
    \scalebox{0.85}{
    \color{black}
 \begin{tabular}[t]{p{5.3cm}|p{0.5cm} | p{0.18cm} p{0.18cm} p{0.18cm} p{0.18cm} p{0.18cm} p{0.18cm}  }
    \hline
         \textbf{Category} & \textbf{Total} & \textbf{AA} & \textbf{EF} & \textbf{HC} & \textbf{IM} & \textbf{DS} & \textbf{SS} \\ \hline\hline  
         \textsc{Information Sharing and Awareness} & \textbf{70} & \textbf{10} & \textbf{10} & \textbf{11} & \textbf{20} & \textbf{7} & \textbf{12} \\ \hline
         \textit{Robot Data Availability} & 10& 2 & 3 & 2  & 1  & 2 & -  \\
         \hline
         \textit{System Status Visibility}   &40 & 6  & 7  & 5  & 14 & 8  & -  \\ \hline
         \textit{Robot Mode Change Notification} & 6 &- & - & - & 1  & 2  & 3   \\ \hline
          \textit{Human Status Visibility} &14 & 2  & - & 4  & 4  & - & 4  \\ \hline \hline
  
          \textsc{Control and Coordination}   &\textbf{62} & \textbf{6}  & \textbf{9}  & \textbf{11} & \textbf{17} & \textbf{17} & \textbf{2}   \\ \hline
          \textit{System Initiation} &3 & - & - & - & 2  & 1  & -  \\ \hline
          \textit{Robot System Configuration}  &15 & 1  & 1  & 6  & 4  & 1  & 2  \\ \hline
          \textit{Responsibility Delineation} & 16 & 1  & 4  & 4  & 3  & 4  & - \\ \hline
          \textit{Multi-Agent Coordination} &8 & - & 3  & - & 5  & - & -  \\ \hline 
         \textit{Range of Control (Levels of Automation)} & 19 &4  & 1  & - & 3  & 11 & -  \\ \hline\hline

        \textsc{Communication and Decision Support} & \textbf{131} & \textbf{24} & \textbf{3}  & \textbf{32} & \textbf{40} & \textbf{8}  & \textbf{24} \\ \hline
           \textit{Decision Support} &22 & 9  & 1  & 7  & 4  & - & 1   \\ \hline
         \textit{Decision Transparency} &23 & 8  & - & 1  & 4  & 4  & 6   \\ \hline
       \textit{Decision Aid Limitation} &0 & - & - & - & - & - & - \\ \hline
          \textit{Interface Design}  &64 & 7  & 1  & 18 & 17 & 4  & 17  \\ \hline
        \rowcolor[HTML]{FAF0DC}\textit{Other}  &21 & - & - & 6 & 15 & - & -  \\ \hline \hline

    \textsc{Safety and Risk Mitigation}  & \textbf{60} & \textbf{3}  & \textbf{8}  & \textbf{1}  & \textbf{43} & \textbf{4}  & \textbf{1}  \\ \hline
    \textit{Environmental Factors} &40 & - & 1  & - & 39 & - & -\\ \hline
    \textit{Safety State Preservation} &7 & 1  & 4  & - & - & 2  & -  \\ \hline
    \textit{Robot Safe Mode} &8 & 1  & 3  & - & 3  & 1  & -  \\ \hline
    \textit{Failure Recovery} &1 & - & - & - & - & - & 1    \\ \hline
    \textit{Override and Shut-Down Capabilities}  &4 & 1  & - & 1  & 1  & 1  & -   \\ \hline \hline

\textsc{Team Performance Monitoring} & \textbf{21} & \textbf{5}  & \textbf{1}  & \textbf{6}  & \textbf{6} & \textbf{3}  & \textbf{-} \\ \hline

\textit{Human Performance Monitoring}  &15 & 5  & - & 6  & 3  & 1  & -  \\ \hline 
\rowcolor[HTML]{FAF0DC}\textit{Other}  &6 & - & 1 & - & 3 & 2  & -  \\ \hline \hline

\textsc{Team Capabilities and Authority}   & \textbf{104} &\textbf{13} & \textbf{30} & \textbf{14} & \textbf{19} & \textbf{14} & \textbf{14} \\ \hline

\textit{System Capabilities}  &78 & 12 & 17 & 13 & 19 & 11 & 6   \\ \hline
\textit{Human Operational Constraints} &17 & 1  & 13 & 1  & - & 1  & 1  \\ \hline
\rowcolor[HTML]{FAF0DC} \textit{Other}  &9 & - & - & - & - & 2 & 7   \\ 
\hline \hline
         
\textbf{Total}
  & \textbf{448} &\textbf{61} & \textbf{61} & \textbf{76} & \textbf{164}
  & \textbf{58} & \textbf{62}  \\
  \hline \hline
\end{tabular}}
\caption{\new{Distribution of 448 utility analysis requirements across taxonomy categories and domains. \textit{Other} subcategory entries indicate candidates for taxonomy extensions. \textit{AA = Aerospace \& Aviation, EF = Emergency \& First Response, HC = Healthcare, IM = Industrial Manufacturing, DS = Defense \& Security, SS = Social \& Service.}}}
\label{tab:utility-distribution}
\end{table}

\section{\new{Using the Taxonomy in Practice}}
\label{sec:discussion}

\new{The taxonomy is intended for both \textbf{RE and HRT researchers} working on autonomous and AI-enabled systems, and \textbf{systems engineering practitioners} in safety-critical domains 
who must produce HRT specifications. For both audiences, we position the taxonomy with respect to the RE process described in standards such as ISO/IEC/IEEE 29148~\cite{ISO29148:2018}.  The taxonomy targets the early stages of this process, i.e., requirements elicitation and specification review, where structural coverage of teamwork concerns is most likely to be missed. It is not intended to replace later stages such as detailed requirement formalization, verification, or compliance demonstration, but to provide a structured input to them.} \new{We see the following primary uses:} \smallskip

\noindent\new{\textbf{Elicitation:}} \new{Without a structured reference, requirement types are easily missed. The 21 subcategories provide a guidance list that helps practitioners cover teamwork concerns, with the examples in \S\ref{sec:lowleveltaxonomy} acting as starting points. This could be paired with the interview guide proposed by \cite{mcdermott2018human} which provides leading questions to help with the elicitation process.}\smallskip

\noindent\new{\textbf{Coverage assessment and gap diagnosis:}} \new{A draft requirement specification can be mapped onto the taxonomy to identify which categories are well-covered and which are sparse. As shown in \S\ref{sec:utilitydemonstration}, several subcategories are consistently under-represented across two independently assembled corpora. Practitioners can use the taxonomy as a diagnostic instrument to flag likely blind spots in their own specifications.}\smallskip

\noindent\new{\textbf{Cross-domain comparison:}} \new{The taxonomy allows requirement profiles to be compared across domains, which is relevant when the same robot platform is deployed across multiple contexts. For example, Boston Dynamics' Spot has been used in Industrial Manufacturing, Defense \& Security, and Healthcare~\cite{li2026comprehensive}. Each of these domains has a different requirement profile: Industrial Manufacturing concentrates on \category{Safety and Risk Mitigation}, Healthcare on \category{Communication and Decision Support} while Defense \& Security focuses primarily on \category{Control and Coordination} and \category{Team Capabilities and Authority}. The taxonomy helps practitioners identify which requirements transfer across these settings, which need domain-specific revision, and where gaps are likely to appear.}  




\section{Threats to Validity}
\label{sec:threats}
We endeavored to be comprehensive throughout the development of our taxonomy but recognize several threats to validity and limitations. To begin, our \new{construction} corpus was obtained via a structured literature review (\S\ref{sec:methodology}). Literature reviews are subject to \textit{coverage error}, meaning that we may have missed relevant literature that could have been revealed if we had used different search terms. We mitigated this by starting our search with well-established literature and applying snowballing, in addition to systematic keyword searches. \new{Further, although we found ``Human-Machine Teaming'' resources (e.g., \cite{mcdermott2018human}), we did not include this as a search term as we wanted to focus on robotics, rather than broader team structures such as cognitive assistants. Including it might have provided additional resources for developing the taxonomy, but it could have reduced its specificity to robotics, which was a key objective.} \textit{Source selection bias} could be present since we focused on English-language sources and Western standards \new{and our validation corpus drew primarily from the HRI conference. Different venues or search keywords may have surfaced additional requirement types.} 

While the expert evaluation serves as a key mechanism for mitigating coverage-related threats and refining the taxonomy, it is itself subject to limitations. Our panel of experts, though deliberately recruited from diverse backgrounds, 
may not be fully representative of the breadth of HRT expertise. Nevertheless, convergent feedback between all experts provides confidence in the taxonomy's broader applicability \new{which is further examined by our utility demonstration}.  

There is a wealth of standards in the aerospace domain, since it is global and highly-regulated, a dominance of aerospace-related results \new{in our construction corpus} threatens \textit{domain coverage}. However, rather than undermining the taxonomy, this reflects the current state of the field, i.e., aerospace and industrial robotics represent some of the most mature and documented areas of HRT, making them an appropriate foundation for an initial taxonomy. We explicitly recognize that emerging domains may surface additional requirement types, and \new{our utility demonstration broadened this to include requirements from diverse domains (see Table \ref{tab:utility-distribution}).} 

Finally, the process of categorizing requirements into a taxonomy involves interpretive judgment. To mitigate \textit{construct validity} threats, the taxonomy was developed iteratively across multiple researchers and refined through expert review, ensuring that classification decisions were not made unilaterally. All requirements \new{in the construction corpus} were successfully classified without needing a residual ``other'' category, which is an indicator of structural completeness with respect to the construction corpus. \new{Our utility demonstration identified a small number of requirements that fell into an 'other' category. The taxonomy will thus be extended to incorporate and evaluate these new subcategories in future work. } 



\section{Conclusions and Future Work}
\label{sec:conclude}
This paper presents a two-level taxonomy of HRT requirements based on 361 requirements from standards and the literature. \new{The taxonomy was developed with the aim to support requirement elicitation, coverage assessment, and cross-domain traceability. }
\new{ We evaluate our taxonomy through an expert evaluation, and a utility demonstration using an independently-constructed corpus of 448 requirements.}

Future work will comprise broader validation across diverse human-robot team configurations, including large-scale industrial case studies. \new{We will also examine whether taxonomy categories exhibit domain-specific precedence. For example, safety may take priority even when information sharing is limited such as in  \cite{jia2025more}. 
Finally, we will investigate the extent to which taxonomy categories can be formalized to support verifiable requirement  specification patterns (e.g., \cite{vazquez2024robotics,etumi2025space}).}  



\noindent\textbf{Data Availability:} Our corpora 
are publicly available in~\cite{teamworktaxonomy2024}.

\noindent\textbf{Acknowledgments:} We are grateful to Amber Drinkwater, Laura Hoang, Lukman Irshad, Caroline Jay, Federico Tavella, and Hannah Walsh for insightful discussions and helpful feedback. We also thank the anonymous reviewers, whose constructive comments substantially strengthened the paper. This work was supported by VeTSS and the Royal Academy of Engineering through a Research Fellowship and a Chair in Emerging Technologies.  This work was also supported by the Centre for Robotic Autonomy in Demanding and Long Lasting Environments (CRADLE) under EPSRC grant EP/X02489X/1. Anastasia was supported by NASA contract 80ARC020D0010.

\bibliographystyle{abbrv}
\bibliography{references}

\end{document}